\documentstyle[11pt,bospasp,twoside,epsf]{article}
\def\edcomment#1{\iffalse\marginpar{\raggedright\sl#1\/}\else\relax\fi}
\reversemarginpar

\begin{document}
\title{Bars and the connection between dark and visible matter}
 \author{E. Athanassoula}
\affil{Observatoire, 2 place Le Verrier, 13248
 Marseille cedex 04, France}

\begin{abstract}
Isolated barred galaxies evolve by redistributing their internal
angular momentum, which is emitted mainly at the inner disc resonances
and absorbed mainly at the resonances in the outer disc and the
halo. This causes the bar to grow stronger and its pattern speed to
decrease with time. A massive, responsive halo enhances this
process. I show correlations and trends between the angular momentum
absorbed by the halo and the bar strength, pattern speed and
morphology. It is thus possible to explain why some disc galaxies are strongly
barred, while others have no bar, or only a short bar or an oval. 
In some cases, a bar is found also in the halo component. This ``halo
bar''
is triaxial, but more prolate-like, is shorter than the disc bar and
rotates with roughly the same pattern speed. I finally discuss whether bars
can modify the density cusps found in cosmological CDM simulations of
dark matter haloes.
\end{abstract}

\section{Introduction}

It is impossible to observe dark matter directly, but its existence
and a number of its properties can be deduced from its effects on
other, visible, galactic components. Thus, properties of the disc and of its
substructure can, if correctly interpreted, give us clues on the
properties of dark matter haloes. Here I will discuss the connection
between bars and dark matter, and what information the former can give us on
the latter.

\section{Angular momentum exchange. Analytical results}

The growth of bars in isolated disc galaxies is governed by the exchange of
angular momentum between different parts of the galaxy. To understand
this better it is best to start with the pioneering paper of
Lynden-Bell \& Kalnajs (1972; LBK). Using linear analytic theory,
these authors showed that it is mainly material at resonance that gains
or loses angular momentum. Material at the inner Lindblad
resonance (ILR) will lose angular momentum, while material
at corotation (CR) and the outer Lindblad resonance (OLR) will
absorb it. Since the spiral/bar within CR is a negative angular momentum
perturbation, feeding it with angular momentum will damp it, while
taking angular momentum from it will excite it. Following in their
footsteps, Athanassoula (2003; A03) added a halo (or, more generally, a
spheroidal component) and applied the results to the case of bars. 
Provided the halo distribution function is a function of
the energy only, halo material at {\it all} resonances will gain
angular momentum. This result should be generalisable to other
distribution functions, provided energy is the main functional
dependence and an appropriate perturbation expansion can be used. 
Both for the disc 
and for the halo, there is more angular momentum lost/gained at a
given resonance if the density is higher there and if the resonant
material is colder and thus more responsive (A03).

So, if a disc galaxy has no halo, or if the latter cannot
participate in the angular momentum exchange, the inner disc will
emit angular momentum, which will be absorbed by the outer disc. On
the other hand, if a halo is present and responsive, then it also will
absorb angular momentum. So more angular momentum can be extracted
from the inner parts in the presence of a responsive dark matter halo
and the bar will grow stronger than in its absence (A03). This 
explains the `paradox' that bars in halo dominated disc galaxies may grow
stronger than in disc dominated cases (Athanassoula \& Misiriotis
2002; AM02). 

Tremaine \& Weinberg (1984) and Weinberg (1985) used nonlinear theory
to follow the effect of angular momentum exchange on the slowdown of
the bar. They find that, as the bar loses angular momentum, it will slow
down, as expected. These works, put together with the analytical studies
discussed above, lead to the firm prediction that bars 
should become stronger and rotate slower in the
presence of massive and responsive haloes. 
 
\section{Angular momentum exchange. Simulation results}

\begin{figure}[h]
\plotone{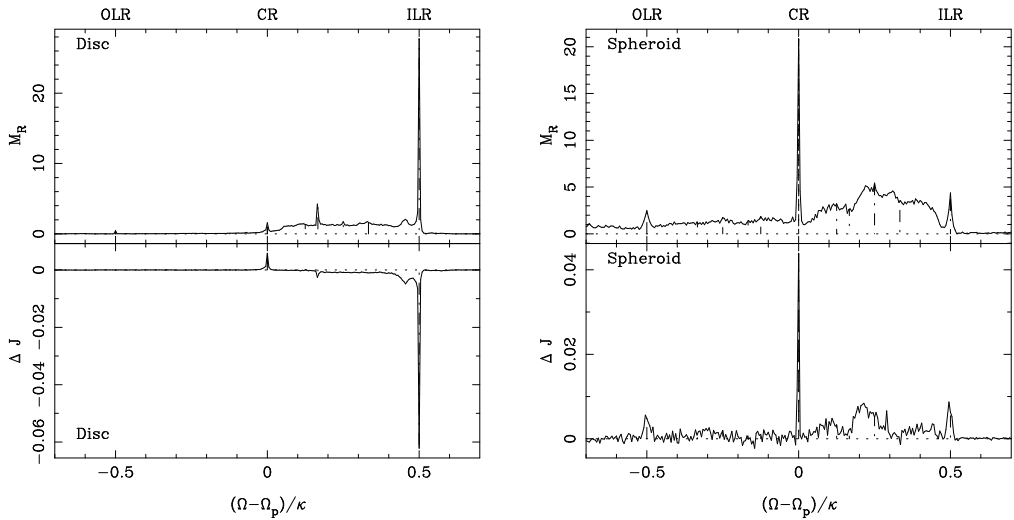}
\caption{Resonances and angular momentum exchange.
The upper panels give the mass per unit frequency ratio as 
a function of that ratio at a time towards the end of the
simulation. The lower panels give $\Delta J$, the angular 
momentum gained, or lost, by the particles between two times, both
taken after the bar has formed, 
again as a function of the frequency ratio. The left 
panels correspond to the disc and the right ones to the spheroid. 
The vertical dash-dotted lines give the positions of the main resonances. }
\label{fig:resonances}
\end{figure}

In Athanassoula (2002; A02) and A03 I tested the above analytical results
with the help of $N$-body simulations and I will follow the same path
here. The first point to test is that there are indeed stars at
(near-) resonance in the halo and that they absorb angular momentum, as
predicted by the analytical results. This is shown in
Fig.~\ref{fig:resonances}, which has been obtained as described 
in A02 and A03. The upper panels plot the mass per unit 
frequency ratio, $M_R$, as a function of that frequency ratio, 
namely $(\Omega - \Omega_p)/\kappa$. Here $\Omega$ is the angular
frequency, $\kappa$ is the epicyclic frequency and $\Omega_p$ is the bar
pattern speed. These panels show that the distribution is not at all
uniform, but has strong peaks at the resonances. This is
true both for the disc, as expected, but also for the halo, as shown
initially in A02. The lower panels show the change of angular momentum
with time (A03),
again as a function of the frequency ratio. Note that the theoretical
predictions of LBK, as well as those of A03, are well confirmed by the
above results, since these show that a large fraction of both the disc
and the halo particles are at resonance, and that they emit/absorb 
angular momentum, as predicted.

\begin{figure}
\plotone{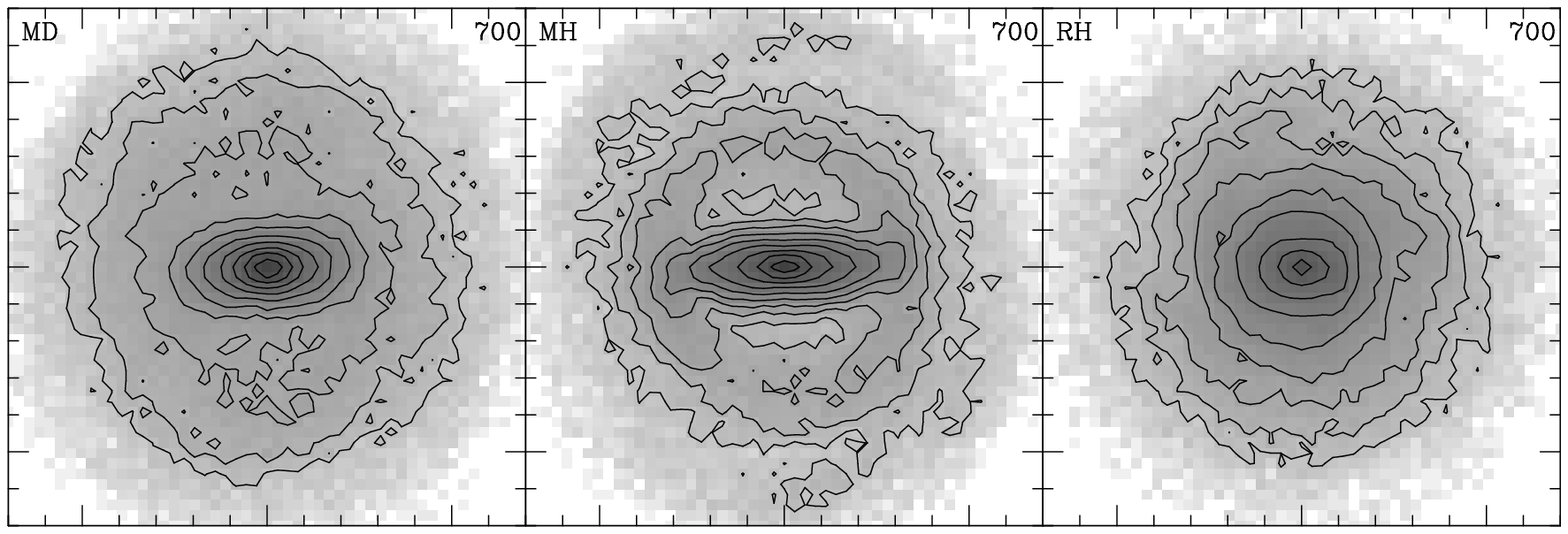}
\caption{Effect of the halo on bar evolution : Face-on views of 
the results of three simulations with different halo components.}
\label{fig:rigid}
\end{figure}

I next check how important the effect of the resonances is.
This can be seen in Fig.~\ref{fig:rigid}, where I compare
the face-on view of the results of three simulations. Initially, the
leftmost is disc dominated in the inner parts, or, in the notation of
AM02 and A03, it is of MD-type. On the contrary, the simulation shown
in the middle panel has initially a large relative halo mass in the
inner parts, i.e. it is of MH-type. Finally, the 
simulation at the right is initially exactly the same as the previous
one, except that the halo is rigid (RH). In this case, the halo is only
present as a rigid potential, so it can not participate in the angular momentum
exchange. The difference between the middle
and right panels is indeed stunning! Instead of the strong bar of the
MH case, the RH case displays a short oval, confined to the inner
parts. Furthermore, comparing the left and middle
panels one can clearly see that the strongest bar has grown in the most
halo dominated environment. These comparisons, and many others in
AM02, A02 and A03,
argue strongly for the role of resonances in bar evolution.

\begin{figure}
\plotone{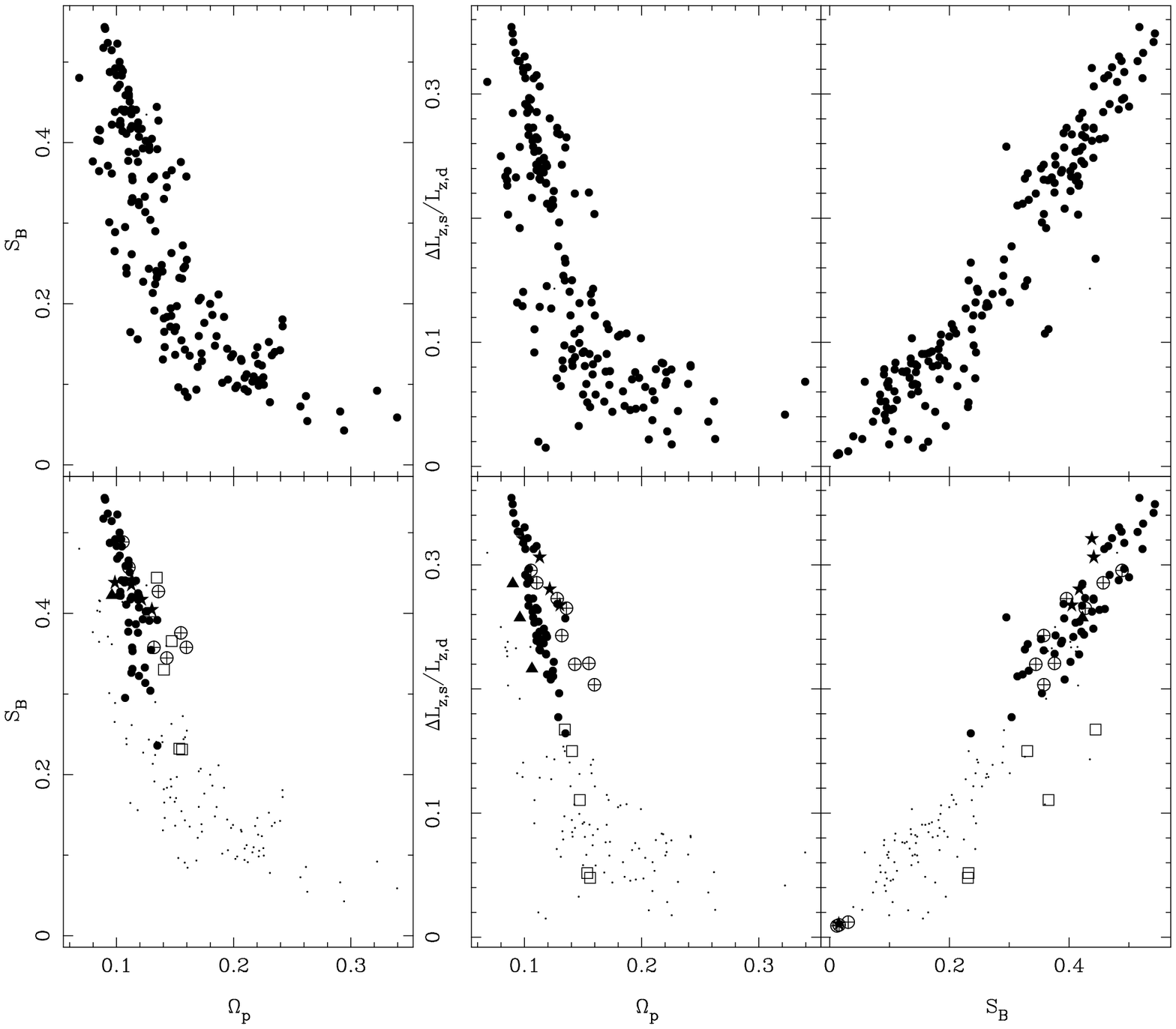}
\caption{Relations between the strength and the pattern speed of the bar
(left panels), between the fraction of the total angular momentum in 
the spheroid and the bar pattern speed (middle panels) 
and between the same quantity and the bar strength (right 
panels). Each symbol represents the result of one simulation (A03). In the
lower panels, results of simulations with a small initial 
halo core radius are given with a large symbol, while those with a 
large core radius with a dot. In particular, simulations with bulges are
marked with a $\oplus$, while filled triangles, filled circles and filled
stars denote simulations with different initial core radius, in order
of increasing 
concentration (see A03). Finally, open squares are for simulations
similar to those denoted by filled
circles, but which have initially very hot discs.} 
\label{fig:correl}
\end{figure}

Theory predicts that both the amount of mass at resonance and its
velocity dispersion should influence the amount of angular momentum
exchanged, and therefore the bar strength and the pattern speed
decrease. A03, with the help of appropriate sequences of simulations,
showed this to be indeed true in the case of $N$-body
simulations. Thus, the increase of the bar strength and the decrease of its
pattern speed are jointly set by three factors : the relative amount of halo
mass at (near-) resonance, how hot the disc is at the
resonances, and how hot the halo is at the resonances. Any one on its
own is not sufficient to determine the outcome, since all three can
limit the amount of angular momentum exchanged.

Since the amount of angular momentum exchanged is a determining
factor, one expects to find correlations between this quantity
and the bar evolution. Unfortunately, the angular momentum exchanged is
a difficult quantity to measure, so in Fig.~\ref{fig:correl}, as in
A03, I use instead the total angular momentum gained by the halo,
relative to the total angular momentum initially in the system. This,
however, is a good approximation 
of the total angular momentum exchanged only in cases when the outer disc
absorbs only little angular momentum. Even so,
Fig.~\ref{fig:correl} shows, for a very large number of simulations, a
clear 
correlation between the relative amount of angular momentum absorbed
by the halo and the bar strength, as well as a trend between it and the bar
pattern speed. Furthermore, if I limit myself to simulations where the
halo is the main absorber (lower panels), then the latter trend
becomes a very strong correlation (middle lower 
panel). These results are basically the same as those shown in A03,
except that I have added here a few more simulations, which have been
run since that paper was written. 

How much are the above correlations dependent on the particular type of
initial conditions used? It is reasonable to assume that initial
conditions having different density and velocity profiles
for the disc, bulge and halo component will also give such
correlations, since these correlations are the reflection of the
physics underlying bar evolution. However, it could well be that the
regression lines (or loci defined by the trend) are somewhat differently
located (shifted) on the corresponding planes. That can be properly
tested only by repeating this type of work with different initial
models; a rather daunting task seen the very large number of simulations 
necessary. Yet a
few clues exist already. For example, from the middle lower panel one
can see that the regression line is somewhat shifted to the right for 
simulations with a more concentrated spheroidal. Similarly, the right
bottom panel shows that the regression line is somewhat lowered for
simulations with initially hotter discs. The differences, however, are
rather small, and 
one can conclude that more angular momentum exchange will
lead to stronger bars that rotate slower. 

Berentzen et al. (2003) present a study where the
bar is driven by a companion in a disc in which a previous bar had
been destroyed by gas inflow. They find good 
agreement of the results of their simulations with those presented
here on the ($S_B$, $\Omega_p$) plane, even though the initial 
models and particularly the problem at hand are very different. 

\begin{figure}
\plotone{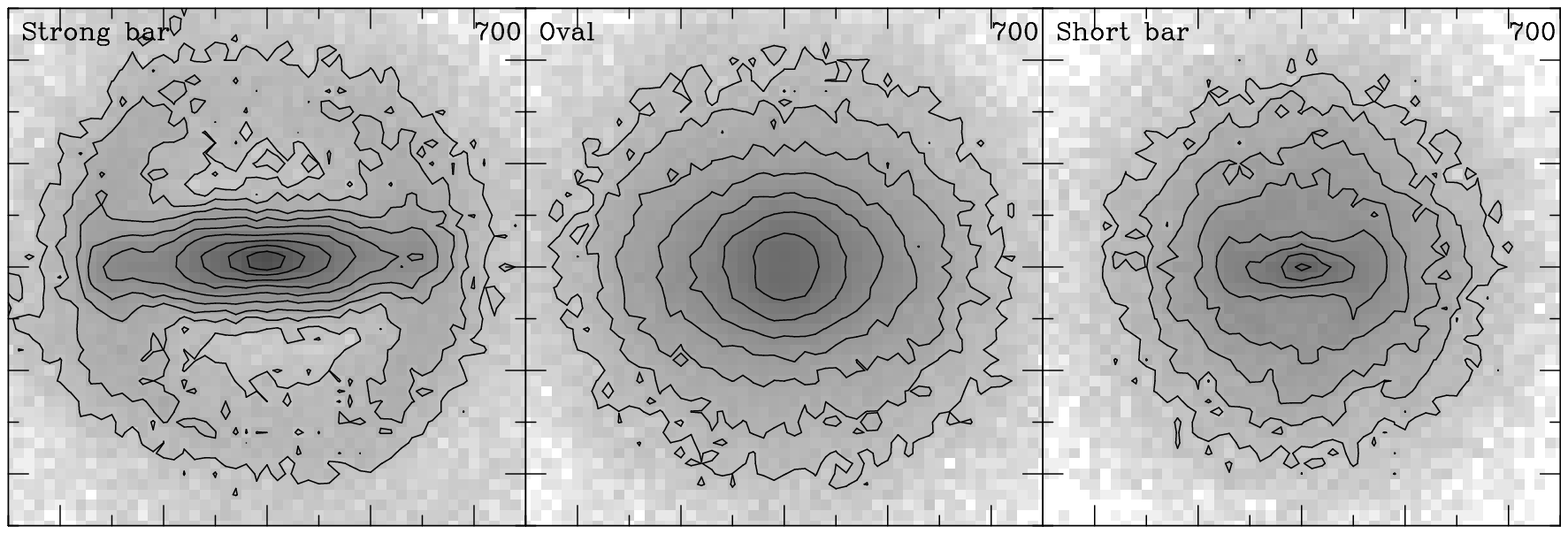}
\caption{Effect of the amount of angular momentum exchanged on the 
morphology of the galaxy : Face-on view of the results of three 
simulations in which different amounts of angular momentum have 
been exchanged between different parts of the galaxy.}
\label{fig:shapes}
\end{figure}

How does the amount of angular momentum exchanged influence the
morphology of the bar? Fig.~\ref{fig:shapes} shows the results of
three simulations. In the one shown in the left panel a large amount
of angular momentum was exchanged during the evolution. It has a 
strong bar, as is the case for all such simulations.
The middle and right panels show examples from simulations
where little angular momentum has been exchanged. In the middle one,
the disc is initially hot and the halo is relatively cool and  concentrated. 
Thus the outer disc can not
contribute much to the angular momentum exchange and it is the
halo that absorbs most of the angular momentum. The result is an oval,
which 
extends to large radii, in order to maximize the emitting region. In
the last example (right panel) the halo is very hot, so it can not
absorb much angular 
momentum, while the disc is initially relatively cool. Thus the bar has to  
stay short, in order for its CR and OLR to be in regions of relatively 
high density, so as to contribute an efficient angular momentum sink.
To summarize, one can say that cases where a lot of angular
momentum has been exchanged always display strong bars. On the other
hand, in cases where only little angular momentum has been exchanged, the
bar can either be short, or longer but more like an oval, depending on
whether it is the disc or the halo that is the main absorber.  

\section{Why is there such a wide variety of bar strengths in observed
  galaxies, and why are some galaxies non-barred ?}

Observed bars come in a wide variety of strengths, ranging from 
very strong bars, like NGC 4608 (Gadotti \& de Souza 2003) or NGC 1365,
to short bars, like our own Galaxy, or ovals, like in NGC
1566. Furthermore, a few discs have no bar at all (Gr\"osbol, Pompei \&
Patsis 2002). How does that compare with the $N$-body results? 

Haloes were initially thought to stabilise discs (e.g. Ostriker \&
Peebles 1973). More recent work (AM02, A02, A03) shows that this
statement is not necessarily correct, since 
haloes can stimulate bar growth by taking angular momentum from the
inner disc. Nevertheless, bars forming in disc galaxies with a
substantial halo take 
{\it longer} to form than bars forming in galaxies with less halo,
so that, in a sense, a halo can be considered as having a stabilising
influence, although the bars that grow in a more massive and
responsive halo may finally become stronger. 
A03 showed that this can happen for a wide range of halo
to disc mass ratios, provided of course the halo is responsive. This,
however, will not necessarily extend to very low relative disc masses (A03).
Simulations to test this limit would be very CPU intensive,
since they would necessitate a very large number of particles, and
also because the bar growth would be exceedingly slow. Moreover, if the
bar grows in time scales much longer than the Hubble time, the problem is
of little astronomical interest and the corresponding galaxies can be
considered bar-stable. 

Haloes can have a further stabilising influence either if
their mass distribution is such 
that not much material is at resonance with the bar, or, more likely,
if the resonant halo material is very hot and thus can not absorb much
angular momentum. Unfortunately, not much is known about the velocity
distribution within the halo component in real galaxies.
 
Similarly, a bar forming in a hot disc will take longer to grow than
in a cold one (Athanassoula \& Sellwood 1986, A03). In very hot discs,
when the bar eventually forms, it has the form of an oval (see
Fig.~\ref{fig:shapes}). 

Thus, with the help of $N$-body simulations, we can account for the 
large variety of bar
strengths observed in real galaxies. As discussed also in the previous
section, this will be determined by the total amount of  angular 
momentum exchanged and also by the specific amount by which each of
the partners (inner 
disc, outer disc, halo, bulge) enters in the exchange. In this picture,
galaxies with no bars should have a very low relative disc mass
and/or a very hot disc and/or a rather unresponsive halo, so that the bar
takes very long to grow and so that the halo can not help its secular evolution.  

This picture is not complete. There is still the question of black
holes, and other `extreme' central concentrations, which should 
have a stabilising influence. The
interplay between this and the responsiveness of the disc and halo
will be discussed elsewhere.

\section{A bar in the halo}

In cases with a sufficiently strong bar in the disc component, the
halo does not stay axisymmetric, but shows also a bar-like or oval
deformation. An example can be seen in Fig. 2 of Holley-Bockelmann, 
Weinberg \& Katz (2003; HBWK). I found such structures also in 
my own simulations and I call them, for simplicity, halo
bars. Preliminary results show that, in cases with a strong disc bar,
the   
halo bar is triaxial, but prolate like, with axial ratios of the
order of 0.7 or 0.8 in the inner parts, and becomes more spherical as
the radius increases. The length of the halo bar increases with time, but
always stays considerably shorter than that of the disc bar. It is 
roughly aligned with the disc
bar at all times (at least within the measuring errors), i.e. it turns
with roughly the same pattern speed. This means that it is a slow bar. The
bisymmetric component in the halo extends well beyond the end of the halo
bar, and there it trails behind the disc bar, much as seen in Fig. 2
of HBWK. The 
properties of these halo bars will be discussed in detail elsewhere 
(Athanassoula, in prep.).  
 
\section{Do bars flatten or steepen central halo cusps? }

Current cosmological
CDM simulations predict that dark matter haloes are cuspy (e.g. White,
this volume), although there is still
no agreement reached about the value of the inner
slope. On the other hand, observations have shown that, at least in a number of
cases where the data are of sufficiently good quality, 
haloes have cores (e.g. Bosma, or de Blok, this volume). If haloes
were indeed formed with a cusp and now have a core, then some
mechanism during galaxy formation and/or evolution should be
responsible for this change. Several have been so far proposed, of
which one involves a bar. Indeed, Hernquist \& Weinberg (1992) and
Weinberg \& Katz (2002), based on analytical calculations and on 
$N$-body simulations of a cusped halo containing a rigid bar 
and no disc, proposed that the bar could flatten the cusp by
giving angular momentum to the halo. This, however, was not generally
accepted, since the fully
self-consistent simulations of Sellwood (2003; S03) and Valenzuela \& Klypin
(2003; VK) showed the contrary, i.e. that during the evolution the halo
radial profile showed a small, yet clear, steepening. These authors 
attribute the flattening found by the previous simulations to
the fact that they do not include a live disc, and thus 
neglect the slow gradual change of the resonant positions due to the
slowdown of the bar, and the disc responsiveness. This explanation did
not, in turn, satisfy HBWK, whose fully 
self-consistent simulations showed a clear flattening of the halo cusp
(HBWK) and thus argued that self-consistency, or the lack of it, could
not be the factor determining why some of the previous
simulations showed a flattening and others a steepening. HBWK instead 
argue that the cusp steepening found by S03 and VK
is spurious, and due to the
fact that their simulations are noisy and thus do not
describe the resonances sufficiently accurately. Numerical
noise could indeed, if sufficiently strong, prevent the halo resonances from
absorbing the correct amount of angular momentum and thus lead to a
wrong evolution. Other numerical shortcomings, however,
e.g. leading to a wrong coupling between the planar and vertical
resonances, could also influence the results, producing a
spurious flattening or steepening. 

Faced with the difficult task of reviewing this unsatisfactory state
of affairs, I turned to my own simulations (AM02, A02, A03) to see
whether they could provide any clues. They are neither grid based, nor
use SCF, and thus could provide an independent view. They were
not specifically designed to tackle this problem, so 
the initial halo has a core (albeit sometimes small). They can, nevertheless,
lead to a number of 
insightful conclusions, which I will discuss briefly below. In
particular, my study of the resonances allows me to single out and to
follow the resonant stars. I could thus make sure that they cover
the phase space adequately. Furthermore, in A02 I showed that there is a large
trapping of the particles at resonances, which shows that they are not
unduly knocked off their trajectories by noise during the
evolution. Thus, my simulations fulfill all the necessary conditions set
by HBWK to perform the test at hand.

In order to check whether the density profile in the inner parts 
flattens or steepens, I simply measured the amount of mass in
concentric spherical shells\footnote{Although very straightforward, this
method is not the most appropriate for the problem at hand, because,
as discussed in the previous section, the halo isodensities are
triaxial and not spherical. Results based on other methods of
calculating the density will be discussed elsewhere.}. Since the
existence of a bar is necessary for the mechanism to work, I confined
myself to times after the bar had grown. I find that, in
the vast majority of the cases I checked, there 
is a steepening of the halo radial density profile, albeit not large. 
In a couple of cases, however, I noted a very small flattening. 
Since this is even smaller in amplitude than the small steepening found in the 
remaining cases, I was ready to discard it as insignificant, until 
I noticed that it occurred in the more centrally concentrated cases. 
This prompted me to investigate the problem further. 

The analytical predictions are clear and follow directly from what was 
discussed in the previous sections : Halo material at (near-) resonance  
should absorb angular momentum and move to larger cylindrical
radii\footnote{It is important in this problem to distinguish between
  cylindrical and spherical radius.}. For the
case of the ILR, such material is located in the inner parts of the
halo and this should lead to a flattening of the  
cusp. I have already shown in the previous sections that my  
simulations confirm the theoretical predictions about the existence 
of the resonances and the angular momentum exchanged. Showing the
increase of the cylindrical radius of particles on orbits at, or
around, the ILR 
is no an easy task, since they are trapped around periodic orbits of
the $x_1$ tree\footnote{The $x_1$ tree is the 3D orbital analog of the
  2D $x_1$ family and provides the backbone of the 3D bars (Skokos,
  Patsis \& Athanassoula 2002).}, for which it is not always easy to
calculate the frequencies and the time average of the radius. Moreover,
there are relatively few particles at ILR in the models I simulated
(Fig.~\ref{fig:resonances}). On the other hand, it is very easy to see
the increase in cylindrical radius in the case of CR, which does not
suffer from the above drawbacks. 

This, however, is not the complete picture, since there are three
effects working against the previous one, and thus leading to a steepening
of the halo :

\begin{itemize}
\setlength{\itemsep}{0pt plus1pt minus1pt}
\setlength{\topsep}{0pt plus1pt minus1pt}
\setlength{\partopsep}{0pt plus1pt minus1pt}
\item As stressed by AM02, VK and 
  O'Neill \& Dubinski (2003), the radial density profile of the disc
  becomes considerably more centrally concentrated with time. In  
  so doing, the disc material pulls the halo material also inwards, thus 
  causing it to  contract. This leads to an increase of the halo
  radial density profile in the inner parts. 
\item As the halo gains angular momentum it will get flattened towards
  the disc equatorial plane and this will result in a decreasing of the
  spherical radius of the individual particles. As seen in the
  previous section, this happens mainly in the inner parts of the halo
  and will, therefore, lead to a steepening of the cusp.
\item During the evolution the bar becomes stronger and this influences
  the shape of the periodic orbits of the $x_1$ tree and of the
  regular orbits trapped 
  around them. Namely, their axial ratio $a/b$ increases and they 
  approach nearer to the center, so that sometimes the trapped orbits
  can actually cover the central area. This means that, although their
  average cylindrical radius may increase, the central-most area may
  have an increased density.
\end{itemize}

There are thus competing effects : On the one hand the resonant
particles can move to larger cylindrical radii and thus tend to
flatten the cusp. On the other hand, there are other effects, also
linked to the 
angular momentum redistribution, which tend to diminish the spherical
radii of the particles. It is the outcome of the competition between these 
effects that will decide whether the inner density
profile will become steeper or shallower during the evolution. It is
thus not necessarily surprising that HBWK find a flattening of the
cusp, while S03 and VK find a steepening. The final
result will depend on the distribution function of the halo, of which hardly
anything is known, as well as that of the disc. In as far as the
$N$-body simulations are 
concerned, it could also, unfortunately, depend on numerical
effects, which could artificially modify the effect of the resonances.

The application of this mechanism to real galaxies faces, to my mind,
two further serious problems. One is that haloes have
substructure. Although this will not inhibit resonance driven
evolution (Weinberg 2001), it could still
perturb the effect of the resonances. The second, perhaps even more
serious, problem is the presence of gas. All simulations that have so
far studied this mechanism are purely stellar and
have no gas. A gaseous component could increase further the
central concentration of the disc material (e.g. Athanassoula 1992,
Heller \& Shlosman 1994) and thus make its inwards
pull on the halo material yet stronger. It would then become yet more
difficult for the halo (near-) resonant stars to overcome the inwards pull and
achieve a flattening of the cusp.

\end{document}